\documentclass[prd,showpacs,groupedaddress]{revtex4-1}
\usepackage{mathrsfs}
\usepackage{amsmath}
\usepackage{amssymb}
\usepackage{graphicx}
\usepackage{dcolumn}
\usepackage{bm}

\def\PLB{Phys.\,Lett.  B}

\def\PRL{Phys.\,Rev.\,Lett.}

\def\PRD{Phys.\,Rev. D}

\begin{document}

 \title{CP violation in D meson decays: would it be a sign of new physics ?}

\author{Franco Buccella}
\email{franco.buccella@na.infn.it}
\affiliation{INFN, Sezione di Napoli,  via Cintia, 80126 Napoli, Italy} 

\author{Maurizio Lusignoli}
\email{maurizio.lusignoli@roma1.infn.it}

\author{Alessandra Pugliese}
\email{alessandra.pugliese@roma1.infn.it}
\affiliation{Dipartimento di Fisica, Sapienza, Universit\`{a} di Roma, Piazzale Aldo Moro 2, 00185 Roma, Italy \\
INFN, Sezione di Roma, Piazzale Aldo Moro 2, 00185 Roma, Italy}

\author{Pietro Santorelli}
\email{pietro.santorelli@na.infn.it}
\affiliation{Dipartimento di Fisica, Universit\`{a} di Napoli Federico II, via Cintia, 80126 Napoli, Italy \\
INFN, Sezione di Napoli,  via Cintia, 80126 Napoli, Italy} 

\begin{abstract}
Ascribing the large SU(3) violations in the singly Cabibbo suppressed decays
of neutral $D$ mesons to the final state interactions, one gets large strong phase
differences, necessary for substantial direct CP violation. While the absolute value of the CP violating
asymmetries depend on the uncertain strength of the penguin contribution, we predict an asymmetry for the decays into charged pions more than twice as large and having opposite sign with respect to that for charged kaons. 
\end{abstract}


\maketitle

\section{Introduction}
\label{introduction}
The experimental results on CP violation in singly Cabibbo suppressed (SCS) decays of the $D^0$ and ${\bar D}^0$ mesons, larger of the common expectation beforehand, published one year ago \cite{CDF,LHCb1} after the less conclusive results of the beauty factories \cite{Belle, BaBaR} have recently been contradicted by new analyses  by the LHCb Collaboration that gave smaller results and moreover of different signs according to the method used \cite{LHCb2,LHCb3}. Defining the CP violating asymmetry for decay into the final state $f$ as $a(f)= [\Gamma(D^0 \to f)-\Gamma(\bar{D}^0 \to f)] / [\Gamma(D^0 \to f)+\Gamma(\bar{D}^0 \to f)]$, the difference of asymmetries in the decays into charged kaons and charged pions, 
$\Delta_{\rm CP}= a(K^+ K^-) - a(\pi^+ \pi^-)$, has been measured with the contradicting results:
\begin{eqnarray}
\label{exp}
\Delta_{\rm CP} &=& (-0.62 \pm  0.21 \pm 0.10)\% \; (\rm CDF) ,\\
&=& (-0.82 \pm  0.21 \pm 0.11)\% \; (\rm LHCb1) , \\
&=& (-0.87 \pm  0.41 \pm 0.06)\% \; (\rm  Belle), \\
&=& (+0.24 \pm  0.62 \pm 0.26)\% \; (\rm  BaBaR), \\
&=& (-0.34 \pm  0.15 \pm 0.10)\%  \; (\rm LHCb2), \\ 
&=& (+0.49 \pm 0.30 \pm 0.14)\%  \; (\rm LHCb3). 
\end{eqnarray}
A naive weighted average  \cite{Seminar} would give $\Delta_{\rm CP} = (-0.33 \pm 0.12)\%$, also  compatible with a null result.

Be as it may, we address the following question: is a CP violation as large as the first indications \cite{CDF,LHCb1} a sign of new physics, as many authors suggested \cite{NewPhys}, or not \cite{StanMod,Silvestrini}?  We favor the second hypothesis.  An important caveat concerns the rather low mass of the decaying meson, that makes the long distance effects very important, and the calculation from first principles impossible.

Many years ago we \cite{ourold} presented a calculation of the decay  branching ratios of $D$ and $D_{\rm s}$ mesons, based on factorization and a model to account for rescattering effects through nearby resonances. The results were in reasonably good  agreement with the data (at that time), but  predicted CP violation at least one order of magnitude smaller than what was found in \cite{CDF,LHCb1}. The experimental data however did change in the meantime, so we decided to make a new analysis, limiting our consideration to the SCS decays. 

In \cite{ourold} we observed that the large flavor SU(3) violations in the data were mainly due to the rescattering effects (because of the difference in mass of the relevant resonances). Therefore we now assume SU(3) symmetry for the weak decay amplitudes prior to rescattering. Furthermore, we approximate the hamiltonian for $D$ weak decays with its $\Delta U = 1$ part when estimating branching ratios, introducing the $\Delta U = 0$ terms only for the calculation of asymmetries. This is justified by the smallness of the relevant CKM elements, 
$|V_{\rm ub} V_{\rm cb}^*| << |V_{\rm ud(s)} V_{\rm cd(s)}^*|$.

\section{Decay branching ratios}
\label{DecayBR}
The weak effective hamiltonian for SCS charmed particles decays can be written as follows:
\begin{equation}
\label{Hamilt}
\mathcal{H}_{\rm w} = \frac{G_F}{\sqrt 2} V_{ud}\,V_{cd}^*
\;[C_1 Q_1^d+C_2 Q_2^d] \;
+ {{G_F}\over{\sqrt 2}} V_{us}\,V_{cs}^*\;[C_1 Q_1^s+C_2 Q_2^s] \;
- {{G_F}\over{\sqrt 2}} V_{ub}\,V_{cb}^*\;\sum_{i=3}^6\;C_i Q_i + h.c. 
\end{equation}

In equation (\ref{Hamilt}) the $C_i$ are Wilson coefficients that multiply the four--fermion operators  defined as \cite{Gilman} 
\begin{eqnarray}
Q_1^d &=& \bar{u}^{\alpha}\,\gamma_{\mu}(1-\gamma_5)
d_{\beta}\,\bar{d}^{\beta}\,\gamma^{\mu}(1-\gamma_5)\,c_{\alpha}\;, \nonumber \\
Q_2^d &=& \bar{u}^{\alpha}\,\gamma_{\mu}(1-\gamma_5)
d_{\alpha}\,\bar{d}^{\beta}\,\gamma^{\mu}(1-\gamma_5)\,c_{\beta}\;, \nonumber \\
Q_3 &=& \bar{u}^{\alpha}\,\gamma_{\mu}(1-\gamma_5)\,c_{\alpha} 
\sum_q\, \bar{q}^{\beta}\gamma^{\mu}(1-\gamma_5)\,q_{\beta}, \nonumber \\
Q_4 &=& \bar{u}^{\alpha}\,\gamma_{\mu}(1-\gamma_5)\,c_{\beta} 
\sum_q\, \bar{q}^{\beta}\gamma^{\mu}(1-\gamma_5)\,q_{\alpha},  \\
Q_5 &=& \bar{u}^{\alpha}\,\gamma_{\mu}(1-\gamma_5)\,c_{\alpha} 
\sum_q\, \bar{q}^{\beta}\gamma^{\mu}(1+\gamma_5)\,q_{\beta}. \nonumber \\
Q_6 &=& \bar{u}^{\alpha}\,\gamma_{\mu}(1-\gamma_5)\,c_{\beta} 
\sum_q\, \bar{q}^{\beta}\gamma^{\mu}(1+\gamma_5)\,q_{\alpha}. \nonumber
\end{eqnarray}
The operators $Q_1^s$ and $Q_2^s$ are obtained by the replacement $d \to s$ in $Q_1^d$ and $Q_2^d$.

The hamiltonian can be decomposed in two parts, according to the change it induces in the $U$ spin. The dominant part has $\Delta U = 1$ and it is
\begin{eqnarray}
H_{\Delta U = 1} &=& {{G_F}\over{2\,\sqrt 2}} (V_{us}\,V_{cs}^*-V_{ud}\,V_{cd}^*)
[C_1( Q_1^s-Q_1^d)+C_2 (Q_2^s-Q_2^d)] \\
 &\simeq &{{G_F}\over{\sqrt 2}} \sin \theta_C \cos \theta_C 
[C_1( Q_1^s-Q_1^d)+C_2 (Q_2^s-Q_2^d)]. \nonumber
\end{eqnarray}
The remaining part, that using the unitarity of the CKM matrix can be written in the form
\begin{equation}
\label{Uzero}
H_{\Delta U = 0} =  - \; {{G_F}\over{\sqrt 2}} V_{ub}\,V_{cb}^*\;\Big\{\sum_{i=3}^6\;C_i Q_i + {1 \over 2}[C_1( Q_1^s+Q_1^d)+C_2 (Q_2^s+Q_2^d)]\Big\}\;,
 \end{equation}
 may be neglected in the calculation of decay branching ratios (even if necessary for CP violation) given that  $|V_{\rm ub} V_{\rm cb}^*| << \sin \theta_C \cos \theta_C$.
In this approximation, the neutral charmed meson $D^0$ being a $U$-spin singlet,  only two independent amplitudes are needed for $D^0$ SCS  decays into two pseudoscalars belonging to SU(3) octets. 
In fact, there are two independent combinations of $S$-wave states having $U$=1:
\begin{eqnarray}
&&{1\over 2}\Big\{|K^+\,K^-> +|K^-\,K^+>-|\pi^+\,\pi^->-|\pi^-\,\pi^+>\Big\}\; ;  \\
&&{\sqrt{3} \over {2 \sqrt{2}}} \Big\{|\pi^0 \, \pi^0>-|\eta_8\,\eta_8>-{1 \over \sqrt{3}}(|\pi^0\,\eta_8>+
 |\eta_8\,\pi^0>)\Big\}\; ,  \nonumber
 \end{eqnarray}
that may be combined in two states with given trasformation properties under SU(3):
\begin{eqnarray}
\label{8SU3}
|8,U=1> = \frac{\sqrt{3}} {2 \sqrt{5}} &\Big\{&|K^+K^-> +|K^-K^+>-|\pi^+\pi^->-|\pi^-\pi^+>  \\
 &-&  \big[|\pi^0\pi^0>-|\eta_8\eta_8>-{1 \over \sqrt{3}}(|\pi^0\eta_8>+
 |\eta_8\pi^0>)\big]\Big\}, \nonumber \\
 \label{27SU3}
 |27,U=1> = \frac {1} {\sqrt{10}} &\Big\{&|K^+K^-> +|K^-K^+>-|\pi^+\pi^->-|\pi^-\pi^+>  \\
  &+& {3 \over 2}\,\big[|\pi^0\pi^0>-|\eta_8\eta_8>-{1 \over \sqrt{3}}(|\pi^0\eta_8>+
 |\eta_8\pi^0>)\big]\Big\} . \nonumber
 \end{eqnarray}
Another independent amplitude would appear considering decays to states involving an SU(3) singlet. In order to keep the number of parameters to a minimum we disregard decays to states containing the singlet $\eta_1$ meson, and therefore we are also neglecting the mixing between the physical states $\eta$ and $\eta$'. That's why the prediction of the model concerning decays to states involving the $\eta$ meson are to be taken {\it cum grano salis}.

Note that eqs.(\ref{8SU3},\ref{27SU3}) would imply no decay to neutral kaons ($K^0 \, \bar{K}^0$) and that decays to charged pions should be more frequent than to charged kaons because of the larger phase space, given the equal and opposite amplitudes. Both predictions are in violent disagreement with experiment. The large SU(3) violations have been much discussed in the literature, a general first order analysis was done many years ago \cite{Savage} and in recent works \cite{NewPhys,StanMod} its relevance to CP violation has been stressed.
 
In our model the necessary SU(3) breaking is determined by the final state interactions,  described as the effect of resonances in the scattering of the final particles. Assuming no exotic resonances belonging to the 27 representation, the possible resonances have SU(3) and isospin quantum numbers $(8, I=1)$, $(8,I=0)$ and $(1,I=0)$.  Moreover, the two states with $I=0$ can be mixed, yielding two resonances:
\begin{eqnarray}
|f_0> &=& \sin \phi  \;\; |8,I=0> + \cos \phi \;\; |1,I=0> , \\
|f'_0> &=& - \cos \phi  \;\; |8,I=0> + \sin  \phi \;\; |1,I=0>. 
\end{eqnarray}
The mixing angle $\phi$ and the strong phases $\delta_0$, $\delta'_0$ and $\delta_1$ are our model parameters, together with the two independent weak decay amplitudes. The strong phases should be related to the mass $M_i$ and total width $\Gamma_i$ of the corresponding resonance through the relation $\tan \delta_i = \Gamma_i / [ 2 (M_i - M_{D^0})]$, however the data on these scalar resonances are sparse and do not allow a clean determination of the phases. 
One plausible hypothesis is that the phase $\delta_1 \sim \pi/2$, since the isovector partner of the scalar resonance $K^*_0 (1950)$ should have a mass close to the $D^0$ mass, as it follows deriving it from an equispacing formula \cite{ourold}.
Note also that we are putting to zero the small phase $\delta_{27}$, so that the $\delta_i$ parameters actually correspond to the differences with respect to the phase in the non resonant channel.

The two independent weak amplitudes can be related to the commonly used diagrammatic amplitudes $T$ and $C$ (color connected and color suppressed respectively) \cite{Chau} in the following way:
\begin{eqnarray}
\label{amplitudes}
A_8(U=1) &\propto& T - \frac {2}{3} \;C \;,  \\
 A_{27}(U=1) &\propto& T + C . \nonumber
 \end{eqnarray}
 Note that in our approach, differently from other authors, both $T$ and $C$ are real numbers, 
 the strong phases being introduced as effects of rescattering. 
 As an example, we consider the decay in charged pions and kaons including the SU(3) violations:
 \begin{eqnarray}
 \label {pipi}
A(D^0 \to \pi^+ \pi^-) & = &   \Big(T-{2 \over 3}C\Big)
 \left\{
  - \frac{3}{10} \left(e^{\imath \delta_0} +   e^{\imath \delta^\prime_0}\right) \right. \\ 
 & + & \left. \left( - \frac{3}{10} \cos(2 \phi) + \frac{3}{4\sqrt{10}} \sin(2 \phi)\right) 
\left(e^{\imath \delta^\prime_0} - e^{\imath \delta_0}\right)
 \right\}\nonumber\\
 & - &\Big(T+C\Big)\;\;
 \frac{2}{5}\; ,\nonumber\\
\label{KK}
A(D^0\to K^+K^-) & = & \Big(T-{2 \over 3}C\Big)
 \left\{
 \frac{3}{20} \left(e^{\imath \delta_0} +   e^{\imath \delta^\prime_0}\right) \right.
\\ 
 & + & \left. \left( \frac{3}{20} \cos(2 \phi) + \frac{3}{4\sqrt{10}} \sin(2 \phi)\right) 
\left(e^{\imath \delta^\prime_0} - e^{\imath \delta_0}\right)
 +  \frac{3}{10} e^{\imath \delta_1}\right\}\nonumber\\
 & + & \Big(T+C\Big)\;\;
 \frac{2}{5}\; .  \nonumber
 \end{eqnarray}
The limit of exact flavor SU(3) would correspond to  $\sin(\phi)$ = 1, $\delta_0$ = $\delta_1$. In this limit the amplitudes do not depend on 
$\delta^\prime_0$ (since in the approximation of keeping only the
$\Delta U = 1$ hamiltonian the $D^0$ meson does not couple to the singlet
state) they are of opposite sign and equal respectively to:
\begin{equation}
A[D^0 \to \pi^+ \pi^-(K^+K^-)] \;\;\to\;\; \mp \left[  \Big(T-{2 \over 3}C\Big)\; \frac{3}{5}\,
e^{\imath \delta_0} + \Big(T+C\Big)\; \frac{2}{5} \right]\; .
\end{equation}
The expressions for the remaining amplitudes are given in the Appendix.

As it can be seen from the above equations, the SU(3) breaking corrections do not change the part of the amplitudes belonging to the 27 representation, but only the octet part, that also acquires a singlet component. Therefore, in our model the SU(3) breaking hamiltonian transforms as a triplet under SU(3), completely analogous to the simplifying hypothesis put forward in \cite{Savage}, first suggested in \cite{Wise}. However, the number of parameters in our model is six, three of which describe the SU(3) symmetry breaking, while in \cite{Savage} the symmetry breaking parameters are four.

We note that the experimental results for the decays of neutral and charged $D$ mesons in a pion pair when analyzed in terms of amplitudes of given isospin $A_2$ and $A_0$, defined by 
${\mathcal A}(D^0 \to \pi^+ \pi^-) = (\sqrt 2\; A_0 - A_2) / \sqrt 6$, 
give \cite{Silvestrini}:
\begin{eqnarray}
\label{isospin}
|A_2| &=& (3.08 \pm 0.08) \; 10^{-7} \; {\rm GeV} \, ,\\
|A_0| &=& (7.6 \pm 0.1)\; 10^{-7} \; {\rm GeV} \, ,\nonumber \\
{\rm arg}(A_2 / A_0) &=& \pm(93 \pm 3)^{\circ} \,. \nonumber
 \end{eqnarray}
On the contrary, the presence of two independent amplitudes with isospin 1 in the $K \bar K$ channels does not allow a determination of the amplitudes from their decay branching ratios.

We found a good description of the experimental data for the rates with the following set 
of parameters (the upper or lower signs should be taken simultaneously):  
\begin{eqnarray}
\label{param}
C \; / \;  T &=& \;-\; 0.529 \, , \\
\sin (2 \phi) = 0.701\,,&\qquad& \cos (2 \phi) = 0.713 \; , \nonumber \\
\sin \delta_0 = \pm \, 0.529 \,,&\qquad& \cos \delta_0 = -\,0.848 \; , \nonumber \\
\sin \delta'_0 = \pm \, 0.794 \,,&\qquad& \cos \delta'_0 = 0.608 \; , \nonumber \\
\sin \delta_1 = \pm \, 0.992 \,,&\qquad& \cos \delta_1 = 0.126 \; . \nonumber 
\end{eqnarray}

In fact, using them we obtain the following results for the ratios of decay rates:
\begin{eqnarray}
\label{result1}
\frac{\Gamma(D^0 \to K_{\rm S}K_{\rm S})}{\Gamma(D^0 \to K^+K^-)} &=& 0.0429\; , \\
\frac{\Gamma(D^0 \to \pi^+ \pi^-)}{\Gamma(D^0 \to K^+K^-)} &=& 0.354 \; ,\nonumber \\
\frac{\Gamma(D^0 \to \pi^0 \pi^0)}{\Gamma(D^0 \to K^+K^-)} &=& 0.202 \; ,\nonumber 
\end{eqnarray}
to be compared to the experimental values \cite{PDG}: 0.043$\pm$0.010, 0.354$\pm$0.010, 
0.202$\pm$0.013, respectively.
Moreover, the ratio of the moduli of the two pion isospin amplitudes is $|A_2 / A_0| = 0.40$ and its phase is  $ \mp 87.2^{\circ}$, in fair agreement with the experimental results reported in eq.(\ref{isospin}).
The result for the absolute values of the branching ratios, obtained using the experimental lifetime,  agree within 20\% with the values obtained using naive factorization (that may be derived in the $\pi^+\,\pi^-$ case from eq. (2.16) of \cite{ourold}). 

It may appear that describing four experimental data (the three ratios in eq.(\ref{result1}) and the analogous ratio for the two pion decay of a $D^+$, or equivalently the relative phase of the two pionic amplitudes with given isospin) with five parameters is trivial. However, 
four of these parameters are angles, and  sines or cosines may only vary between $-1$ and $1$, so that formulae like those given in the Appendix are not capable of describing any number. The result presented in eq.(\ref{param}) has not been obtained with a least squares fit, and not every parameter has been taken as really free. In fact, we required $|\sin(\delta_1)| \simeq  1$ (as already said above) and  $C \; / \;  T \sim - \; 0.5$, similar to the results of our old fits  \cite{ourold}. 

Finally, we note that identifying the $\eta$ meson with $\eta_8$ the branching ratios to final states  would come out $\Gamma(D^0 \to \pi^0 \eta)\;/\;\Gamma(D^0 \to K^+K^-) = 0.216$ and 
$\Gamma(D^0 \to \eta \eta)\;/\;\Gamma(D^0 \to K^+K^-) = 0.250$, to be compared to the experimental values (0.172$\pm$0.018, 0.422$\pm$0.051) respectively. Also in this case, the rescattering effects are helpful in allowing a decay rate to $\eta \eta$ larger than to $\pi^0\pi^0$, albeit to an insufficient level, in spite of the phase space difference.

\section{CP asymmetries}
\label{CPasy}
A nonzero direct CP asymmetry is present only when the decay amplitude is a sum of two  amplitudes with different weak phases and having also two different strong phases. If the amplitude for $D$ decay is
$${\mathcal A} = A \; e^{\imath \delta_A} + B \; e^{\imath \delta_B}\:,$$
the CP conjugate amplitude would be
$${ \bar {\mathcal A}} = A^* \; e^{\imath \delta_A} + B^* \; e^{\imath \delta_B}\: ,$$
and the CP asymmetry is:
\begin{eqnarray}
\label{asym}
a_{\rm CP} &=& {{|{\mathcal A}|^2-|{ \bar {\mathcal A}}|^2} \over {|{\mathcal A}|^2+
|{ \bar {\mathcal A}}|^2}} \\
&=& {{2\; \Im (A^*\,B)\; \sin (\delta_A - \delta_B)} \over
{|A|^2 + |B|^2 + 2 \; \Re (A^*\,B)\; \cos (\delta_A - \delta_B)}}.
\end{eqnarray}

The second amplitude $B$ is provided in our case by the matrix elements of the $\Delta U = 0$ hamiltonian, eq.(\ref{Uzero}), that contains both $Q_{1(2)}$ and "penguin" operators. In this case, there are three independent symmetric states of two pseudoscalar mesons:
\begin{eqnarray}
\label{Unull}
&&{1\over 2}\Big\{|K^+\,K^-> +|K^-\,K^+>+|\pi^+\,\pi^->+|\pi^-\,\pi^+>\Big\}\; ;   \\
&&{1 \over 4} \Big\{3\,|\pi^0 \, \pi^0>+|\eta_8\,\eta_8>+\sqrt{3}\,(|\pi^0\,\eta_8>+
 |\eta_8\,\pi^0>)\Big\}\;  ; \nonumber \\
 &&{1 \over \sqrt{3}} \Big\{\frac{1}{4}|\pi^0 \, \pi^0>+\frac{3}{4}\,|\eta_8\,\eta_8>-\frac{\sqrt{3}}{4}\,(|\pi^0\,\eta_8>+
 |\eta_8\,\pi^0>)+  |K^0\,\bar{K^0}>+|\bar{K^0}\,K^0>\Big\}\; , \nonumber
 \end{eqnarray}
that give rise to three amplitudes transforming as 27, 8 and 1 under SU(3) (for the $Q_{1(2)}$ part) and to two amplitudes transforming as 8 and 1 (for the penguin part). In the framework of quark diagrams (and neglecting annihilation) the third state in eq.(\ref{Unull}) decouples, both for penguins and for the other terms. Moreover, the $\Delta I = 1/2$ property of the penguin selects one combination of the first two states. Taking into account that now 
also the singlet components of the resonances couple to the $D^0$ meson state, after rescattering the relevant amplitudes become:
 \begin{eqnarray}
 \label {pipiUz}
 B(D^0 \to \pi^+ \pi^-) & = &  \left(P+\frac{T'}{2}\right)
 \left\{
   \frac{1}{2} \left(e^{\imath \delta_0} +   e^{\imath \delta^\prime_0}\right)
  +  \left( - \frac{1}{6} \cos(2 \phi) - \frac{7}{4\sqrt{10}} \sin(2 \phi)\right) 
 \left(e^{\imath \delta^\prime_0} - e^{\imath \delta_0}\right) \right\} \\
&+& \left( T' + C'  \right) \; \left\{ \frac{3}{20} -
 \frac{3}{40}\; \left(e^{\imath \delta_0} +   e^{\imath \delta^\prime_0}\right)  \right. \nonumber\\
&+& \left. \Big[ {1 \over 120} \, \cos (2 \phi) + {1\over {4 \sqrt {10}}}\,\sin (2 \phi) \Big] \;
 \left(e^{\imath \delta^\prime_0} - e^{\imath \delta_0}\right) \right\} \;,\nonumber \\
\label{KKUz}
 B(D^0\to K^+K^-) & = &  \left(P+\frac{T'}{2}\right)
 \left\{
 \frac{1}{4} \left(e^{\imath \delta_0} +   e^{\imath \delta^\prime_0}\right)  
  +   \left( -\frac{5}{12} \cos(2 \phi) + \frac{1}{4\sqrt{10}} \sin(2 \phi)\right) 
\left(e^{\imath \delta^\prime_0} - e^{\imath \delta_0}\right)
 +  \frac{1}{2} e^{\imath \delta_1}\right\}\\
 & + & 
 \left( T'+C' \right) \left\{\frac{3}{20}\; - \frac{1}{40}\; \left(e^{\imath \delta_0} +   
 e^{\imath \delta^\prime_0}\right) + \frac{7}{120} \, \cos (2 \phi)
 \left(e^{\imath \delta^\prime_0} - e^{\imath \delta_0}\right)  - \frac{1}{10}\, e^{\imath \delta_1} \right\}.\nonumber
 \end{eqnarray}
 
 In eqs.(\ref{pipiUz},\ref{KKUz}) $P$ is the contribution of the ``penguin'' diagram, while the other parameters $T'$ and $C'$ are related (in the framework of quark diagrams without  annihilations) to $T$ and $C$ by the relations
 \begin{equation}
 \label{Dparam}
 T'= - \; T \; \frac{V_{ub}\, V_{cb}^*}{\sin\theta_C \cos \theta_C}\quad {\rm and} \quad
 C'= - \; C \; \frac{V_{ub}\, V_{cb}^*}{\sin\theta_C \cos \theta_C}\;.
 \end{equation}
We note that if $T'+C'=0$ the terms containing these amplitudes have the same structure of the penguin term, and that therefore could be reabsorbed in the uncertainty of the penguin contribution. 
In our phase convention the amplitudes $T$ and $C$ are real, while $T'$, $C'$ and $P$ are complex, having the phase $\pi - \gamma = (111 \pm 4)^\circ$ \cite{PDG,UTfit}.

The numerical value of the ratios $|T'/T|$  and $|C'/C|$ being (6.6$\pm$0.9) $\cdot 10^{-4}$, they would result in a CP asymmetry of this order. A large asymmetry may only be due to the penguin contribution. 
We recall that the penguin diagrams were introduced as a possible explanation of the ``octet enhancement'' by Shifman, Vainshtein and Zacharov \cite{SVZ} many years ago. A large matrix element for these operators could successfully describe both the kaon and the hyperon non--leptonic decays. There has not been a general consensus on this approach, and in particular a recent lattice calculation \cite{Sachr} seems to indicate a different origin for the $\Delta I= 1/2$ dominance in kaon decays. 

Neglecting the contribution of the terms containing $T'$ and $C'$,  the amplitude for a particular decay channel, say $K^+ K^-$, using the equations (\ref{KK},\ref{KKUz}) can be written as
$${\mathcal A}(K^+K^-) \simeq T\; f_T(\delta_i,\phi,C/T)\;+\; P\;f_P(\delta_i,\phi) \;,$$
and  equation (\ref{asym}) gives
\begin{equation}
\label{asym2}
a_{CP}(K^+ K^-) \simeq \frac {2 \; T \; \Im (P) \; \Im(f_T\, f_P^*)}{T^2\; |f_T|^2 + ...}
\end{equation}
where we neglected terms of order $|P|/T$ in the denominator, an approximation already made in the calculation of the decay rates.

Inserting in the relevant formulae the parameter values previously determined from the branching ratios and choosing the lower signs in eq.(\ref{param}), the CP asymmetries for decays in charged mesons turn out to be

\begin{eqnarray}
\label {result}
a_{CP}(K^+ K^-) &=& \frac{\Im (P)}{T} \cdot (+ 1.469) \; , \\
a_{CP}(\pi^+ \pi^-) &=& \frac{\Im (P)}{T} \cdot (- 3.362) \; . \nonumber
\end{eqnarray}
The sign would be opposite if one chooses instead the upper signs in eq.(\ref{param}).
Our choice is  suggested by the fact that apparently the resonance $f_0$(1710) - that has a lower mass - prefers to decay in a pair of kaons \cite{PDG} and should therefore be identified with $f'_0$.

We also report the prediction for CP asymmetries for decays in final states with neutral mesons, although it will probably be difficult to test them by experiment:
\begin{eqnarray}
\label {predict}
a_{CP}(K^0 {\bar K}^0) &=& \frac{\Im (P)}{T} \cdot (-1.217) \; , \\
a_{CP}(\pi^0 \pi^0) &=& \frac{\Im (P)}{T} \cdot (-1.668) \; . \nonumber
\end{eqnarray}

We note that our parameters predict an asymmetry in the decay to charged pions that is of opposite sign with respect to the asymmetry for decays to charged kaons, and more than twice as large.
Assuming instead equal values for the phases $\delta_0, \delta'_0$ and $\delta_1$, the asymmetries  
would be equal and opposite, but of considerable less magnitude (even for a maximal strong phase). Therefore, the SU(3) breaking in rescattering favors, in a sense, a larger $\Delta_{CP}$. 
Taking into account the CKM elements entering in the definition of $T$ and $P$, one has 
\begin{equation} 
 \frac{\Im (P)}{T} = \frac{|V_{ub}\,V_{cb}|}{\sin \theta_C \cos \theta_C} \sin \gamma 
 \frac{<K^+\,K^-| \; \sum_{i=3}^6\;C_i Q_i + {1 \over 2}[C_1\{Q_1^s+Q_1^d\}+C_2 \{Q_2^s+Q_2^d\}]
  \;| D^0>}
 {<K^+\,K^-|\,C_1( Q_1^s-Q_1^d)+C_2 (Q_2^s-Q_2^d)\,|D^0>} 
 = 6.3\,10^{-4} \kappa \; , 
\end{equation}
 where the notation $<K^+\,K^-|\;\{Q_i\}\;| D^0>$ indicates the matrix element evaluated with a penguin contraction of the operator. One obtains therefore:
\begin{equation} 
 \Delta_{\rm CP} =  3.03 \; 10^{-3} \kappa \; .
 \end{equation}
 A value of $\kappa$ around three gives asymmetries at the percent level.  Concerning the sign of  $\Delta_{\rm CP}$, we note that if one uses factorization $\kappa$ would be negative and 
$\Delta_{\rm CP}$ would therefore be negative, in agreement with the majority of experimental results.
We note however that if one uses factorization a considerably smaller value for $\kappa$ would be expected, due to the smallness of the Wilson coefficients of QCD penguin operators.

Let us compare this result to what has been found in \cite{Silvestrini}, where an analysis of the bounds imposed by unitarity on the final state interactions of the isospin zero amplitudes was pursued, both in a two--channel and in a three--channel situation. We note that the enhancement factor $\kappa$ required is similar to what was found there in the three channel case, and that, in our SU(3) based scheme, the channels are in fact three (1, 8, 27).

\section{Conclusion}
In this paper, we analyzed the singly Cabibbo suppressed  decays of the neutral $D$ mesons in the framework of a model that ascribes all of the large SU(3) violations to final state interactions. The values of the strong phases are therefore large and in principle suitable to predict consistent CP violations in the decay amplitudes. We were able to give an accurate description of decay branching ratios and of the isospin structure of the amplitudes for pionic decays. 

The experimental situation regarding the CP violating asymmetries is at present rather confused, but we think anyhow of interest to have shown that large asymmetries can be obtained, considering the uncertainties of long distance contributions and with some stretching of the parameters, even without invoking New Physics.
The final CP asymmetries depend on the value of the "penguin" matrix element, and a rather large value would be needed to obtain asymmetries as large as in \cite{CDF,LHCb1,Belle}. We recall that large "penguin" contributions were also suggested to reproduce rates and isospin structure of the decays of $K$ mesons and hyperons \cite{SVZ}, although it is not evident that the analogy can be pursued \cite{Wise}. While the absolute value of the CP violating
asymmetries cannot be safely predicted, we obtain an asymmetry for the decays into charged pions more than twice as large and having opposite sign with respect to that for charged kaons.

\appendix
\section{}
\label{App}
In this Appendix we collect all the formulae for the decay amplitudes, including those already given above. All decay amplitudes are given by a sum of two terms, ${\mathcal A}= A + B$. 
The rates are obtained in the usual way from the squared moduli of the amplitudes, with a factor 1/2 when the two particles in the final state are identical. 
The main contribution to their square modulus, determining the decay rates, is given below as amplitude $A$:
\begin{eqnarray}
\label{AK+K-}
A(D^0\to K^+K^-) & = & \Big(T-{2 \over 3}C\Big)
 \left\{
 \frac{3}{20} \left(e^{\imath \delta_0} +   e^{\imath \delta^\prime_0}\right) \right. \\ 
 & + & \left. \left( \frac{3}{20} \cos(2 \phi) + \frac{3}{4\sqrt{10}} \sin(2 \phi)\right) 
\left(e^{\imath \delta^\prime_0} - e^{\imath \delta_0}\right)
 +  \frac{3}{10} e^{\imath \delta_1}\right\}\nonumber\\
 & + & \Big(T+C\Big)\;\;
 \frac{2}{5}\; , \nonumber 
 \end{eqnarray}
 \begin{eqnarray}
 \label{AK0}
 A( D^0 \to K^0 {\bar K}^0) & = & \Big(T-{2 \over 3}C\Big)
 \left\{
 \frac{3}{20} \left(e^{\imath \delta_0} +   e^{\imath \delta^\prime_0}\right) \right. \\
  & + & \left. \left( \frac{3}{20} \cos(2 \phi) + \frac{3}{4\sqrt{10}} \sin(2 \phi)\right) 
\left(e^{\imath \delta^\prime_0} - e^{\imath \delta_0}\right)
 -  \frac{3}{10} e^{\imath \delta_1} \right\} \; ,\nonumber \\
 \label {Apipi}
A(D^0 \to \pi^+ \pi^-) & = &   \Big(T-{2 \over 3}C\Big)
 \left\{
  - \frac{3}{10} \left(e^{\imath \delta_0} +   e^{\imath \delta^\prime_0}\right) \right. \\ 
 & + & \left. \left( - \frac{3}{10} \cos(2 \phi) + \frac{3}{4\sqrt{10}} \sin(2 \phi)\right) 
\left(e^{\imath \delta^\prime_0} - e^{\imath \delta_0}\right)
 \right\} \nonumber \\
 & - &\Big(T+C\Big)\;\;
 \frac{2}{5}\; ,\nonumber \\
\label {Api0}
A(D^0 \to \pi^0 \pi^0) & = &   \Big(T-{2 \over 3}C\Big)
 \left\{
  - \frac{3}{10} \left(e^{\imath \delta_0} +   e^{\imath \delta^\prime_0}\right) \right. \\ 
 & + & \left. \left( - \frac{3}{10} \cos(2 \phi) + \frac{3}{4\sqrt{10}} \sin(2 \phi)\right) 
\left(e^{\imath \delta^\prime_0} - e^{\imath \delta_0}\right)
 \right\}\nonumber \\
  & + &\Big(T+C\Big)\;\;
 \frac{3}{5}\; ,\nonumber \\
 \label {Aetapi0}
A(D^0 \to \eta_8 \pi^0)  &=&   \Big(T-{2 \over 3}C\Big) \frac{\sqrt{3}}{5}\;e^{\imath \delta_1}
- \Big(T+C\Big)\;\;
 \frac{\sqrt{3}}{5}\; ,  \\
 \label {Aetaeta}
A(D^0 \to \eta_8 \eta_8) & = &   \Big(T-{2 \over 3}C\Big)
 \left\{
   \frac{3}{10} \left(e^{\imath \delta_0} +   e^{\imath \delta^\prime_0}\right) \right. \\ 
 & + & \left. \left(  \frac{3}{10} \cos(2 \phi) + \frac{3}{4\sqrt{10}} \sin(2 \phi)\right) 
\left(e^{\imath \delta^\prime_0} - e^{\imath \delta_0}\right)
 \right\}\nonumber \\
  & - &\Big(T+C\Big)\;\;
 \frac{3}{5}\; .\nonumber 
 \end{eqnarray}
 
  The remaining parts of the decay amplitudes (with different weak phase), indicated by B, are :
 \begin{eqnarray}
 \label{BKK}
 B(D^0\to K^+K^-) & = &  \left(P+\frac{T'}{2}\right)
 \left\{
 \frac{1}{4} \left(e^{\imath \delta_0} +   e^{\imath \delta^\prime_0}\right)  
  +   \left( -\frac{5}{12} \cos(2 \phi) + \frac{1}{4\sqrt{10}} \sin(2 \phi)\right) 
\left(e^{\imath \delta^\prime_0} - e^{\imath \delta_0}\right)
 +  \frac{1}{2} e^{\imath \delta_1}\right\}\\
 & + & 
 \left( T'+C' \right) \left\{\frac{3}{20}\; - \frac{1}{40}\; \left(e^{\imath \delta_0} +   
 e^{\imath \delta^\prime_0}\right)  + \frac{7}{120} \, \cos (2 \phi)
 \left(e^{\imath \delta^\prime_0} - e^{\imath \delta_0}\right)  - \frac{1}{10}\, e^{\imath \delta_1} \right\}.\nonumber \\
 \label{BK0}
 B(D^0\to K^0{\bar K}^0) & = &  \left(P+\frac{T'}{2}\right)
 \left\{
 \frac{1}{4} \left(e^{\imath \delta_0} +   e^{\imath \delta^\prime_0}\right)  
  +   \left( -\frac{5}{12} \cos(2 \phi) + \frac{1}{4\sqrt{10}} \sin(2 \phi)\right) 
\left(e^{\imath \delta^\prime_0} - e^{\imath \delta_0}\right)
 -  \frac{1}{2} e^{\imath \delta_1}\right\}\\
 & - & 
 \left( T'+C' \right) \left\{\frac{1}{20}\; + \frac{1}{40}\; \left(e^{\imath \delta_0} +   
 e^{\imath \delta^\prime_0}\right)  - \frac{7}{120} \, \cos (2 \phi)
 \left(e^{\imath \delta^\prime_0} - e^{\imath \delta_0}\right)  - \frac{1}{10}\, e^{\imath \delta_1} \right\}.\nonumber \\
 \label {Bpipi}
 B(D^0 \to \pi^+ \pi^-) & = &   \left(P+\frac{T'}{2}\right)
 \left\{
   \frac{1}{2} \left(e^{\imath \delta_0} +   e^{\imath \delta^\prime_0}\right)
  +  \left( - \frac{1}{6} \cos(2 \phi) - \frac{7}{4\sqrt{10}} \sin(2 \phi)\right) 
\left(e^{\imath \delta^\prime_0} - e^{\imath \delta_0}\right) 
 \right\}  \\
 &+& \left( T' + C'  \right) \; \left\{ \frac{3}{20} -
 \frac{3}{40}\; \left(e^{\imath \delta_0} +   e^{\imath \delta^\prime_0}\right)  \right. \nonumber\\
&+& \left. \Big[ {1 \over 120} \, \cos (2 \phi) + {1\over {4 \sqrt {10}}}\,\sin (2 \phi) \Big] \;
 \left(e^{\imath \delta^\prime_0} - e^{\imath \delta_0}\right) \right\} \;.\nonumber \\
 \label {Bpi0}
 B(D^0 \to \pi^0 \pi^0) & = &   \left(P+\frac{T'}{2}\right)
 \left\{
   \frac{1}{2} \left(e^{\imath \delta_0} +   e^{\imath \delta^\prime_0}\right)
  +  \left( - \frac{1}{6} \cos(2 \phi) - \frac{7}{4\sqrt{10}} \sin(2 \phi)\right) 
\left(e^{\imath \delta^\prime_0} - e^{\imath \delta_0}\right) 
 \right\}  \\
 &-& \left( T' + C'  \right) \; \left\{ \frac{7}{20} +
 \frac{3}{40}\; \left(e^{\imath \delta_0} +   e^{\imath \delta^\prime_0}\right)  \right. \nonumber\\
&-& \left. \Big[ {1 \over 120} \, \cos (2 \phi) + {1\over {4 \sqrt {10}}}\,\sin (2 \phi) \Big] \;
 \left(e^{\imath \delta^\prime_0} - e^{\imath \delta_0}\right) \right\} \;.\nonumber \\
 \label {Betapi0}
B(D^0 \to \eta_8 \pi^0)  &=&   \left(P+\frac{T'}{2}\right)  \frac{1}{\sqrt 3} \;e^{\imath \delta_1} -
\left( T'+C' \right) \frac{\sqrt{3}}{10}\;\left[1+ \frac {2}{3}\,e^{\imath \delta_1} \right] \; . 
 \end{eqnarray}
 \begin{eqnarray}
 \label {Betaeta}
 B(D^0 \to \eta_8 \eta_8) & = &  \left(P+\frac{T'}{2}\right) \;
 \left\{
\frac{1}{6} \left(e^{\imath \delta_0} +   e^{\imath \delta^\prime_0}\right)
  +  \left( - \frac{1}{2} \cos(2 \phi) + \frac{11}{12\sqrt{10}} \sin(2 \phi)\right) 
\left(e^{\imath \delta^\prime_0} - e^{\imath \delta_0}\right) 
 \right\}  \\
 &+& \left( T' + C'  \right) \; \left\{ - \frac{3}{20} -
   \frac{1}{120} \; \left(e^{\imath \delta_0} +   e^{\imath \delta^\prime_0}\right) \right.
\nonumber \\
& + & \left.  \Big[ \frac{3}{40}\; \cos(2\,\phi)-\frac{1}{12 \sqrt{10}}\; \sin (2\,\phi) \Big] 
\left(e^{\imath \delta^\prime_0} - e^{\imath \delta_0}\right) \right\} \, . \nonumber
\end{eqnarray}


\end{document}